\documentclass[preprint,showpacs,preprintnumbers,amsmath,amssymb]{revtex4}

\usepackage{graphicx}
\usepackage{dcolumn}
\usepackage{bm}

\begin{document}
\title{Vector fields in multidimensional cosmology}
\author{Boris E. Meierovich}
\email{meierovich@mail.ru}
\affiliation{P.L.Kapitza Institute for Physical Problems  \\
2 Kosygina str., Moscow 119334, Russia}
\date{\today}
\homepage{http://www.kapitza.ras.ru/people/meierovich/Welcome.html}

\begin{abstract}
Vector fields in the expanding Universe are considered within the multidimensional theory of General Relativity.


Vector fields in general relativity form a three-parametric variety. Our consideration includes the fields with a nonzero covariant divergence. Depending on the relations between the particular parameters and the symmetry of a problem, the vector fields can be longitudinal and/or transverse, ultrarelativistic (i.e. massless) or nonrelativistic (massive), and so on.

The longitudinal and transverse vector fields are considered separately in detail in the background of the de Sitter cosmological metric. In most cases the field equations reduce to Bessel equations, and their temporal evolution is analyzed analytically. The energy-momentum tensor of the most simple  zero-mass longitudinal vector fields enters the Einstein equations as an additive to the cosmological constant. In this case the de Sitter metric is the exact solution of the Einstein equations. Hence, the most simple zero-mass longitudinal vector field pretends to be an adequate tool for macroscopic description of dark energy as a source of the expansion of the Universe at a constant rate. The zero-mass vector field does not vanish in the process of expansion. On the contrary, massive fields vanish with time.
Though their amplitude is falling down, the massive fields make the expansion accelerated.


\end{abstract}

\maketitle

\section{\label{sec:level1}Introduction} 

	According to modern observations the Universe is expanding as a whole, despite the gravitational attraction between material objects. From the standpoint of general relativity the matter curves the space-time, giving rise to mutual attraction between the bodies. The fact of the Universe expansion confirms the assumption that the physical vacuum is not just the absence of the ordinary matter. The existence of the so called `dark energy' and `dark matter', as the source of the Universe expansion, is widely discussed in modern literature  \cite{Wikipedia}. Can we overcome the ''expansion-attraction'' contradiction within the existing theory of general relativity? That is the question.
	
Without the knowledge of nature of vacuum it is still possible to analyze some general macroscopic properties of physical phenomena
 \cite{example}.
It is worth applying the macroscopic approach, keeping in mind that there is no way to reproduce the full scale cosmology in the laboratory experiments.
	
The purpose of this article is to study the behavior of vector fields in the expanding Universe. Utilization of vector fields in general relativity shows undoubtable advantages in comparison with scalar fields and with multiplets of scalar fields. In brane world applications vector fields play the role of the order parameter at phase transitions with spontaneous symmetry breaking \cite{Meier1}. The equations appear to be more simple, while their solutions are more general. The solutions have additional parametric freedom, allowing to forget the fine-tuning problem.

Having in mind that the brane world is a 3-dimensional manyfold in the space with extra dimensions, the number $d_{0}$ of space-like coordinates is left arbitrary. From the point of view of phase transitions the brane world is a topological defect, inevitably accompanying the phase transitions with spontaneous symmetry breaking. Knowing the dependence of temporal evolution of vector fields on the dimension $d_{0}$ of the initially uniform bulk could help finding out if vector fields can initiate the phase transition.
	
The paper is organized as follows. In Section \ref{Section2} we consider a cosmological metric, describing the multidimensional homogeneous and isotropic expansion, caused by the cosmological constant only.
	
General properties of vector fields in the background of an arbitrary metric are presented in Section \ref{Section3}. We get rid of the restriction (\text{\ref{divfi = 0}}), commonly used in flat space-time in order to avoid the troubles of negative contribution to the energy. In general relativity it is worth replacing by a more weak condition of regularity.
	
Particular properties of vector fields in the background of the cosmological metric (\text{\ref{Cosmological metric}}) are analyzed in Section \ref{Section4}. Longitudinal and transverse fields are considered separately in Subsections \ref{Subsection4.1}\ and \ref{Subsection4.2}, respectively. In the background of the de Sitter metric (\ref{de Sitter metric})  the solutions of vector field equations are found analytically in terms of Bessel functions.
	
Analytical analysis of vector fields in the de Sitter metric shows that the boundary between the nonrelativistic (massive) fields and the ultrarelativistic (zero-mass) ones corresponds to extremely low masses. The zero-mass vector field looks adequate for macroscopic description of the dark energy.
The principle difference between them (massive fields vanish in the process of cosmological expansion, while the massless ones don't) approves the separation of dark relativistic substance into dark matter and dark energy.
	
The results are summarized in Section \ref{Summary}.

\section{\label{Section2} Expansion of the Universe driven by the cosmological constant}
	
According to observations the Universe expands, remaining homogeneous and isotropic. Keeping in mind applications to multidimensional cosmology we consider the ($D=1+d_{0}$)-dimensional space-time having the structure $T^{1}E^{d_{0}}$ and the metric
\begin{equation}
ds^{2}=g_{IK}dx^{I}dx^{K}=( dx^{0})^{2}-e^{2F( x^{0}) }\sum_{I=1}^{d_{0}}( dx^{I})^{2}
\label{Cosmological metric}
\end{equation}
depending on only one time-like coordinate $x^{0}=ct.$ The metric tensor $g_{IK}$ is diagonal. The expansion is characterized by a single function $F( x^{0}) ,$ and the rate of expansion (Hubble parameter) is $H=c\frac{dF}{dx^{0}}\equiv cF^{\prime }$. The Ricci tensor is also diagonal:
\begin{equation}
R_{00}=-d_{0}( F^{\prime 2}+F^{\prime \prime }) ,
\label{Ricci R_00=}
\end{equation}
\begin{equation}
R_{II}=e^{2F}( F^{\prime \prime }+d_{0}F^{\prime 2}) ,\qquad I>0.
\label{Ricci R_11=}
\end{equation}
In the absence of matter there are two Einstein equations for one unknown $F:$
\begin{equation}
-(d_{0}-1)F^{\prime \prime }-\frac{1}{2}d_{0}\left(d_{0}-1\right) F^{\prime 2} -\Lambda =0,\qquad (I\neq 0)
\label{F''-...=Lambda}
\end{equation}
\begin{equation}
\frac{1}{2}d_{0}(d_{0}-1) F^{\prime 2} +\Lambda=0 ,\qquad (I= 0) .
\label{...-d_oF''=-Lambda}
\end{equation}
$\Lambda $ is the cosmological constant.
If $d_{0}>1$ the solution
\begin{equation}
 F=\pm \sqrt{-\frac{2\Lambda }{d_{0}(d_{0}-1) }}( x^{0}-x_{0}^{0})
\label{F=...  linear function of x^0}
\end{equation}
exists if the cosmological constant $\Lambda $ is negative,
\begin{equation}
\Lambda <0.
\label{Lambda<0}
\end{equation}
$F( x^{0}) $ is a linear function; $x_{0}^{0}$ is a constant of integration. The multidimensional $( D=1+d_{0}) $\ cosmological metric
\begin{equation}
ds^{2}=( dx^{0})^{2}-e^{\pm \sqrt{-\frac{8\Lambda }{d_{0}(d_{0}-1) }}( x^{0}-x_{0}^{0}) }\sum_{I=1}^{d_{0}}( dx^{I}) ^{2},\quad d_{0}>1
\label{de Sitter metric}
\end{equation}
is regular. This metric describes the evolution of the homogeneous and isotropic Universe connected with the presence of the cosmological constant $\Lambda $. It is either expansion (sign $+$), or contraction (sign $-$). The metric (\ref{de Sitter metric}) is called de Sitter (or anti de Sitter, depending on the sign definition of the Ricci tensor).
	
In the case of expansion (sign $+$)
\begin{equation*}
H=\sqrt{-\frac{2\Lambda c^{2}}{d_{0}(d_{0}-1) }}
\end{equation*}	
is the multidimensional Hubble constant, i.e. the rate of expansion. In the case of three space coordinates
\begin{equation*}
H=\sqrt{-\frac{1}{3}\Lambda c^{2}},\qquad d_{0}=3.
\end{equation*}	
The today's value of the Hubble constant is \cite{Gravitation}
\begin{equation}
H_{\text{today}}=55\text{ km}\cdot \sec ^{-1}\cdot \text{Mpc}^{-1}\approx 1.8\times 10^{-18}\sec ^{-1}.
\label{H today}
\end{equation}	
Available rough estimate for the Hubble constant at a hypothetical stage of inflation is \cite{Wikipedia2}
\begin{equation}
10^{42}\sec ^{-1}>H_{\text{inf}}>10^{36}\sec ^{-1}.
\label{H inf}
\end{equation}

\section{\label{Section3}Vector field in general relativity}
	
In general relativity, the Lagrangian of a vector field $\phi _{I}$ consists of the scalar bilinear combinations of its covariant
derivatives and a scalar potential $V( \phi ^{K}\phi _{K}) $. A bilinear combination of the covariant derivatives is a 4-index tensor
$S_{IKLM}=\phi _{I;K}\phi _{L;M}.$ The most general form of the scalar $S$, formed via contractions of $S_{IKLM}$, is $S=( ag^{IK}g^{LM}+bg^{IL}g^{KM}+cg^{IM}g^{KL}) S_{IKLM},$ where $a,b,$ and $c$ are arbitrary constants. The general form of the Lagrangian of a vector field $\phi _{I}$ is
\begin{equation}
L=a( \phi _{;M}^{M})^{2}+b\phi _{;M}^{L}\phi _{L}^{;M}+c\phi _{;M}^{L}\phi _{;L}^{M}-V( \phi _{M}\phi ^{M}).
\label{Lagrangian}
\end{equation}

 The classification of vector fields $\phi _{I}$ is most convenient in terms of the symmetric $G_{IK}=\frac{1}{2}( \phi _{I;K}+\phi _{K;I}) $ and antisymmetric $F_{IK}=\frac{1}{2}( \phi _{I;K}-\phi _{K;I}) $ parts of the covariant derivatives. The Lagrangian (\ref{Lagrangian}) gets the form
 \begin{equation*}
 L=a(G_{M}^{M})^{2}+(b+c)G_{M}^{L}G_{L}^{M}+(b-c)F_{M}^{L}F_{\text{ }L}^{M}-V(\phi _{M}\phi ^{M}).
 \end{equation*}
 The bilinear combination of antisymmetric derivatives $F_{M}^{L}F_{\text{ }L}^{M}$ is the same as in electrodynamics. It becomes clear in the common notations $A_{I}=\phi _{I}/2,$ $F_{IK}=A_{I;K}-A_{K;I}.$
	
The terms with symmetric covariant derivatives deserve special attention. In applications of the vector fields to elementary particles in flat space-time the divergence $\frac{\partial \phi ^{K}}{\partial x^{K}}$ is artificially set to zero \cite{Bogolubov-Shirkov}:
\begin{equation}\label{div fi = 0}
\frac{\partial \phi ^{K}}{\partial x^{K}}=0.
\end{equation}
This restriction allows to avoid the difficulty of negative contribution to the energy. In the electromagnetic theory it is referred to as Lorentz gauge. However in general relativity (in curved space-time) the energy is not a scalar, and its sign is not invariant against the arbitrary coordinate transformations. Vector fields are also used as the order parameters in phase transitions with spontaneous symmetry breaking \cite{Meier1}. Among the excitation states of physical vacuum there can be those of the vector type. Considering free vector fields in general relativity, it is worth getting rid of the restriction (\ref{div fi = 0}), using instead a more weak condition of regularity.

The covariant field equations
\begin{equation}\label{Covar field eqs}
a\phi _{;K;I}^{K}+b\phi _{I;K}^{;K}+c\phi _{;I;K}^{K}=-V^{\prime }\phi _{I}
\end{equation}
and the energy-momentum tensor
\begin{equation}\label{T_IK= general}
\begin{array}{l}
T_{IK}=-g_{IK}L+2V^{\prime }\phi _{I}\phi_{K}+2ag_{IK}(\phi_{;M}^{M}\phi^{L})_{;L}  +2(b+c)[(G_{IK}\phi^{L})_{;L}-G_{K}^{L}F_{IL}-G_{I}^{L}F_{KL}] \\
+2(b-c)(2F_{\text{ \ }I}^{L}F_{LK}-F_{\text{ \ }K;L}^{L}\phi_{I}-F_{\text{ \ }I;L}^{L}\phi_{K})
\end{array}
\end{equation}
describe the behavior of vector fields in the background of any arbitrary given metric $g_{IK}$ \cite{Meier2}. Here $V^{\prime }\equiv \frac{dV(\phi_{M}\phi^{M})}{d(\phi_{M}\phi^{M})}$. \qquad
	
If the back reaction of the field on the curvature of space-time is essential, then the metric obeys the Einstein equations
\begin{equation}
 R_{IK}-\frac{1}{2}g_{IK}R+\Lambda g_{IK}=\varkappa T_{IK}
 \label{Einstein equations  General}
\end{equation}
with (\ref{T_IK= general})  added to $T_{IK}.$ Here $\varkappa $ is the (multidimensional) gravitational constant. The field equations (\ref{Covar field eqs}) follow from the Einstein equations (\ref{Einstein equations  General}) due to the Bianchi identities. The field equations (\ref{Covar field eqs}) are linear with respect to $\phi$ if the vector field is small, and the terms with the second and higher derivatives of the potential $V\left(\phi_M \phi^M\right)$ can be omitted.

\section{\label{Section4}Vector field in the background of the cosmological metric}

Depending on the sign of the invariant $\phi^{I}\phi_{I}\ $the vector $\phi^{I} $ is either time-like $(\phi^{I}\phi_{I}>0) ,$ or space-like $(\phi^{I}\phi_{I}<0) .$ In general relativity, while all $D$ coordinates are equivalent, one can choose the appropriate coordinate system where $\phi^{I}=0$ either for $I>0,$ or for $I=0.$ But it can not be done if the coordinate system is already chosen in accordance with some other reasons. In the cosmological metric (\ref{Cosmological metric})  the coordinate $x^{0}$ is already specified, and whatever the sign of the scalar $\phi^{I}\phi_{I}$ is, we have to consider $\phi_{I}$ having both space and time components.
	
However, all space coordinates in the metric (\ref{Cosmological metric}) are equivalent, and we can choose the coordinate $x^{1}$ along the space direction of the vector field. Then the vector $\phi_{I}$ has only two nonzero components $\phi_{0},$ and $\phi_{1}.$ All other space components of the vector $\phi_{I}$ are zeros:
\begin{equation}\label{fi_I>1 = 0}
    \phi _{I}=0,\text{ \ } I>1.
\end{equation}
	
The derivatives of the vector field in the equations (\ref{Covar field eqs})  are:
 \begin{equation}\label{fi_;L;I=}
    \phi _{;L;I}^{L} =\frac{\partial }{\partial x^{I}}\left( \frac{\partial \phi _{0}}{\partial x^{0}}+d_{0}F^{\prime }\phi _{0}-e^{-2F}\frac{\partial \phi _{1}}{\partial x^{1}}\right)
 \end{equation}
\begin{equation}\label{fi_I;L=}
\phi _{I;L}^{;L}=\widehat{D}\phi _{I}+2\delta _{I0}F^{\prime }e^{-2F}\frac{\partial \phi _{1}}{\partial x^{1}}+2F^{\prime }\left( \frac{\partial \phi _{0}}{\partial x^{I}}-\frac{\partial \phi _{I}}{\partial x^{0}}\right) -\delta _{I0}d_{0}F^{\prime 2}\phi _{0}-\delta _{I1}\left( d_{0}F^{\prime 2}+F^{\prime \prime }\right) \phi _{1}
\end{equation}
 \begin{equation}\label{fi_;I;L=}
\phi _{;I;L}^{L}=\frac{\partial }{\partial x^{I}}\left( \frac{\partial \phi _{0}}{\partial x^{0}}+d_{0}F^{\prime }\phi _{0}-e^{-2F}\frac{\partial \phi _{1}}{\partial x^{1}}\right) -\delta _{I0}d_{0}\left( F^{\prime 2}+F^{\prime \prime }\right) \phi _{0}-\delta _{I1}\left( d_{0}F^{\prime 2}+F^{\prime \prime }\right) \phi _{1}
 \end{equation}
Here $\widehat{D}$ is the d'Alamber operator:
\begin{equation*}
    \widehat{D}\phi _{I}=\frac{1}{\sqrt{-g}}\frac{\partial }{\partial x^{K}}\left( \sqrt{-g}g^{KL}\frac{\partial \phi _{I}}{\partial x^{L}}\right) =\frac{\partial ^{2}\phi _{I}}{\left( \partial x^{0}\right) ^{2}}+d_{0}F^{\prime }\frac{\partial \phi _{I}}{\partial x^{0}}-e^{-2F}\sum_{K=1}^{d_{0}}\frac{\partial ^{2}\phi _{I}}{\left( \partial x^{K}\right) ^{2}}.
\end{equation*}

The second derivatives (\ref{fi_;I;L=})  and (\ref{fi_;L;I=}) differ from one another by the order of differentiation. In curved space-time
\begin{equation*}
    \phi _{;I;L}^{L}=\phi _{;L;I}^{L}+R_{LI}\phi ^{L}
\end{equation*}
in accordance with (\ref{Ricci R_00=}) and (\ref{Ricci R_11=}).

In view of (\ref{fi_I>1 = 0}) the field equations (\ref{Covar field eqs})  in the background of the cosmological metric (\ref{Cosmological metric}) are:
\begin{equation}\label{Eq for I=0}
\begin{array}{l}
\left( a+c\right) \frac{\partial }{\partial x^{0}}\left( \frac{\partial \phi _{0}}{\partial x^{0}}+d_{0}F^{\prime }\phi _{0}-e^{-2F}\frac{\partial \phi _{1}}{\partial x^{1}}\right) \\
      +b\widehat{D}\phi _{0}+2bF^{\prime }e^{-2F}\frac{\partial \phi _{1}}{\partial x^{1}}-bd_{0}F^{\prime 2}\phi _{0}-cd_{0}\left( F^{\prime 2}+F^{\prime \prime }\right) \phi _{0}=-V^{\prime }\phi _{0},\quad I=0,
\end{array}
\end{equation}
\begin{equation}\label{Eq for I=1}
\begin{array}{l}
\frac{\partial }{\partial x^{1}}\left[ \left( a+c\right) \left( \frac{\partial \phi _{0}}{\partial x^{0}}+d_{0}F^{\prime }\phi _{0}-e^{-2F}\frac{\partial \phi _{1}}{\partial x^{1}}\right) +2bF^{\prime }\phi _{0}\right]  \\
+b\widehat{D}\phi _{1}-2bF^{\prime }\frac{\partial \phi _{1}}{\partial x^{0}}-\left( b+c\right) \left( d_{0}F^{\prime 2}+F^{\prime \prime }\right) \phi _{1}=-V^{\prime }\phi _{1},\quad I=1,
\end{array}
\end{equation}
\begin{equation}\label{Eqs for I>1}
\begin{array}{l}
\frac{\partial }{\partial x^{I}}\left[ \left( a+c\right) \left( \frac{\partial \phi _{0}}{\partial x^{0}}+d_{0}F^{\prime }\phi _{0}-e^{-2F}\frac{\partial \phi _{1}}{\partial x^{1}}\right) +2bF^{\prime }\phi _{0}\right] =0,\quad I>1.
\end{array}
\end{equation}
There are $D$ equations for only two unknowns $\phi _{0}$ and $\phi _{1}.$
	
The space within the cosmological metric (\ref{Cosmological metric}) is invariant against arbitrary translations and rotations. The coordinates $x^{K},$ $K>0,$ are cyclic variables, and the conjugate momenta $p_{K},$ $K>0,$ are integrals of motion. The vector field $\phi _{I}$\ is a superposition of eigenfunctions $\phi _{I\left(\textbf{p}\right) }\sim e^{ip_{K}x^{K}},$ each satisfying the equations (\ref{Eq for I=0}-\ref{Eqs for I>1}). $D-2$ equations (\ref{Eqs for I>1})  take the form (Here and below we omit the index (\textbf{p}))
\begin{equation}\label{p_I[...]=0}
    p_{I}\left[ \left( a+c\right) \left( \frac{\partial \phi _{0}}{\partial x^{0}}+d_{0}F^{\prime }\phi _{0}-e^{-2F}\frac{\partial \phi _{1}}{\partial x^{1}}\right) +2bF^{\prime }\phi _{0}\right] =0,\quad I>1.
\end{equation}
So there are two possibilities. Either the vector field does not depend on the transverse coordinates $x^{I},$ $I>1,$ and all $p_{I}$ in (\ref{p_I[...]=0}) are zeros, or the expression in square brackets is zero.
	
In the first case
\begin{equation}\label{p_I>1=0}
    p_{I}=0,\quad I>1
\end{equation}
all $D-2$ equations (\ref{Eqs for I>1}) are satisfied identically. The function
\begin{equation}\label{Fi(x^0,x^1)}
    \Phi =\left( a+c\right) \left( \frac{\partial \phi _{0}}{\partial x^{0}}+d_{0}F^{\prime }\phi _{0}-e^{-2F}\frac{\partial \phi _{1}}{\partial x^{1}}\right) +2bF^{\prime }\phi _{0}
\end{equation}
depends on only two coordinates $x^{0},x^{1}.$ The vector field is longitudinal: its space component $\phi _{1}$ is directed along and depends upon the same space coordinate $x^{1}$.
	
In the second case
\begin{equation}\label{Fi=0}
    \Phi =0
\end{equation}
the vector field $\phi _{I}$\ can depend not only on $x^{0}$ and $x^{1},$ but also on other space coordinates. The coordinates $x^{0}$ and $x^{1}$ are already specified above in accordance with the symmetry of the problem. The subspace of all other coordinates $x^{K},$ $K>1,$ still remains uniform and isotropic. It is convenient to denote the coordinates of the subspace $x^{K}, $ $K>1$ so that the projection of the momentum $p_{I}$ on the subspace $x^{K>1}$ is directed along the coordinate $x^{2}:$ $p_{2}\neq 0,$ and $p_{K>2}=0.$ Thus $ x^{2}$ is the only coordinate of the subspace $x^{K},$ $K>1,$ that the field $\phi _{I}$ depends on. $\phi _{I}$ is a pure transverse vector field, if it depends only on one space coordinate $x^{2}$ (and time $x^{0},$ naturally). If $\phi_{I}$ depends on two space coordinates $x^{1}$ and $x^{2},$ then it is a combination of a longitudinal and a transverse fields.
	
Relation (\ref{Fi(x^0,x^1)}) allows to split the system (\ref{Eq for I=0},\ref{Eq for I=1}) into two separate equations for $\phi _{0}$ and $\phi _{1}:$
\begin{equation}\label{bDfi_0+...}
    b\widehat{D}\phi _{0}+\left( \frac{4b^{2}}{a+c}+\left( b-c\right) d_{0}\right) F^{\prime 2}\phi _{0}-\left( cd_{0}+2b\right) F^{\prime \prime }\phi _{0}+V^{\prime }\phi _{0}=\frac{2b}{a+c}F^{\prime }\Phi -\frac{\partial \Phi }{\partial x^{0}}
\end{equation}
\begin{equation}\label{b(Dfi_1-...}
    b\left( \widehat{D}\phi _{1}-2F^{\prime }\frac{\partial \phi _{1}}{\partial x^{0}}\right) -\left( b+c\right) \left( d_{0}F^{\prime 2}+F^{\prime \prime }\right) \phi _{1}+V^{\prime }\phi _{1} =-\frac{\partial \Phi }{\partial x^{1}}.
\end{equation}
In case (\ref{p_I>1=0}) the three equations (\ref{Fi(x^0,x^1)},\ref{bDfi_0+...}, and \ref{b(Dfi_1-...}) for three unknowns $\phi _{0},\phi _{1},\Phi $ describe a longitudinal vector field in the background of the cosmological metric (\ref{Cosmological metric}). The set (\ref{bDfi_0+...},\ref{b(Dfi_1-...}) with $\Phi =0$ and $p_{2}\neq 0$ corresponds to a vector field depending on a transverse coordinate. The projections $\phi _{0}$ and $\phi _{1}$ of a transverse vector field are independent from one another.

The variety of vector fields is characterized by three free parameters $a,$ $b,$ and $c.$ Vector fields were considered during the last decade by several authors in connections with the dark sector, see \cite{Dolgov}-\cite{Zuntz} and references there in. In the background of the de Sitter metric (\ref{de Sitter metric}) vector fields were studied each time for particular values of the parameters  $a,$ $b,$ and $c,$ see the Table \ref{Table}.
 \begin{table}\caption{Values of the free parameters of the Lagrangian (\ref{Lagrangian}) that some people used in their analysis of vector fields in the background of the de Sitter metric (\ref{de Sitter metric}).}
  \centering
        $\begin{pmatrix}
      $This paper$ & \textit{a}  & \textit{b}  & \textit{c}\\
      $\cite{Dolgov} 1999$ & 0 &  \eta_{0}/4 &  0 \\
      $\cite{Rubakov} 2000$ & 0  & \eta/8   & 0 \\
      $\cite{Kiselev} 2004$ & 0 &  \xi/4& 0  \\
      $\cite{Armendariz-Picon} 2004$  & 0  & -1/8 & 1/8\\
      $ \cite{Bohmer} 2007$& 0  & -1/2 & 1/2 \\
     $\cite{Koivisto} 2008$  & 0  & -1/8 & 1/8 \\
     $\cite{Jimenez1} 2008$& -1/8  & -1/8 & 1/8  \\
     $\cite{Jimenez2} 2008$& -\eta/8  & -1/8 & 1/8  \\
     $\cite{Jimenez3} 2009$& -\xi/8  & -1/8 & 1/8  \\
     $ \cite{Zuntz} 2010$ & \sim c_{2}  & \sim c_{1}  & \sim c_{3}\\
  \end{pmatrix}$
    \label{Table}
\end{table}


The term with $a\neq 0$  appears in \cite{Jimenez1}-\cite{Jimenez3} in combination with the electromagnetic field. From my point of view the fields with symmetric and antisymmetric covariant derivatives can have different physical nature, and they deserve to be considered separately as well.

In \cite{Zuntz} the Lagrangian involves all three arbitrary constants. However the vector field in \cite{Zuntz} is supposed to have only time-like nonzero component. This assumption is consistent with transverse vector fields, for their space and time projections are independent from one another. But it excludes some longitudinal fields from consideration. In general time-like and space-like components of a longitudinal vector field in the de Sitter metric are mutually bound.

In what follows the analytical solutions are presented for two particular cases. Namely, a longitudinal vector field with  $b=c=0,$ and a transverse vector field with  $a=0,$ \ \ $b=-c$ are considered in detail.

\subsection{\label{Subsection4.1}Longitudinal vector field}

It looks strange, but the most simple case
 \begin{equation}\label{b=c=0}
    a\neq 0,\text{ \ \ }b=c=0
 \end{equation}
was not yet considered (at least, I haven't found it in the literature). In this case the equations (\ref{Eq for I=0}-\ref{Eqs for I>1})  reduce to
\begin{equation}\label{dFi'dx^I=...  I=0,1}
    \frac{\partial \Phi }{\partial x^{I}}=-V^{\prime }\phi _{I},\quad I=0,1
\end{equation}
\begin{equation}\label{Fi=a(...)  case a}
    \Phi =a\left(\frac{\partial \phi_{0}}{\partial x^{0}}+d_{0}F^{\prime }\phi _{0}-e^{-2F}\frac{\partial \phi _{1}}{\partial x^{1}}\right).
\end{equation}
Evidently, these equations follow directly from (\ref{Covar field eqs}). The divergence $\phi _{;K}^{K}=\Phi /a.$ The energy-momentum tensor (\ref{T_IK= general})  reduces to
\begin{equation}\label{T_IK    case a}
    T_{IK}=g_{IK}\left( \Phi ^{2}/a+V\right) +2V^{\prime }\left( \phi _{I}\phi _{K}-g_{IK}\phi _{L}\phi ^{L}\right) .
\end{equation}
	
As usual the scalar
\begin{equation*}
    V^{\prime }(0)/a=m^{2}
\end{equation*}
in the case (\ref{b=c=0})  can be designated as the square of mass of a vector field. The two cases $V^{\prime }=0$ (massless field)\ and $V^{\prime }\neq 0$ (massive field)  are considered separately below.

\subsubsection{\label{subsubsection}Zero mass vector field. Macroscopic description of dark energy}

The equations (\ref{Covar field eqs})  for a massless field
\begin{equation*}
    m=0
\end{equation*}
are simply
\begin{equation}\label{field equation case a}
    \frac{\partial \phi _{;L}^{L}}{\partial x^{I}}=0,\quad b=c=V^{\prime }=0.
\end{equation}
The divergence of the vector field $\phi _{;L}^{L}$ is a constant scalar:
 \begin{equation}\label{fi^L_;L = Fi_0/a}
    \phi _{;L}^{L}=\frac{\partial \phi ^{0}}{\partial x^{0}}+d_{0}F^{\prime }\phi ^{0}+\frac{\partial \phi ^{1}}{\partial x^{1}}=\frac{\Phi _{0}}{a}.
 \end{equation}
 This is one equation for two unknowns $\phi ^{0},$ and $\phi ^{1}.$ The general solution is
 \begin{eqnarray*}
 \phi ^{0} &=&\frac{\partial \Psi }{\partial x^{1}}+C_{0}e^{-d_{0}F\left( x^{0}\right) }\int_{x_{0}^{0}}^{x^{0}}e^{d_{0}F\left( y\right) }dy, \\
 \phi ^{1} &=&-e^{-d_{0}F}\frac{\partial \left( e^{d_{0}F}\Psi \right) }{\partial x^{0}}+C_{1}\left( x^{1}-x_{0}^{1}\right) ,
 \end{eqnarray*}
where $\Psi $ is an arbitrary function of all $D=1+d_{0}$ coordinates, $x_{0}^{0}$ and $x_{0}^{1}$ are constants, and $C_{0}+C_{1}=\Phi _{0}/a$. In the general solution both projections $\phi _{0}$ and $\phi _{1}$ are mutually connected via $\Psi $.
	
The energy-momentum tensor
\begin{equation}\label{T_IK    case a m=0}
   T_{IK}=g_{IK}\left( \Phi _{0}^{2}/a+V_{0}\right) ,\quad b=c=V^{\prime }=0
\end{equation}
acts in the Einstein equations (\ref{Einstein equations General})  as a renormalizing addition to the cosmological constant:
\begin{equation}\label{Einstein eqs b=c=V'=0}
  R_{IK}-\frac{1}{2}g_{IK}R+\widetilde{\Lambda }g_{IK}=0,\quad \widetilde{\Lambda }=\Lambda -\varkappa \left( \Phi _{0}^{2}/a+V_{0}\right) .
\end{equation}
Here $V_{0}$ is the value of the potential $V\left( \phi _{L}\phi ^{L}\right) $ in the case of massless field $\left( V^{\prime }=0\right) $.

The de-Sitter metric (\ref{de Sitter metric}) with $\Lambda $ replaced by $\widetilde{\Lambda }$\ is the self-consistent solution of the Einstein equations (\ref{Einstein eqs b=c=V'=0}) with account of back reaction. The zero mass vector field is not supposed to be small. Available today properties of the so called dark energy (presently unknown form of matter providing the major contribution to the uniform isotropic expansion of the Universe) can be described macroscopically by the zero-mass vector field with a simple Lagrangian
\begin{equation}\label{Lagrangian case a}
   L=a\left( \phi _{;M}^{M}\right) ^{2}-V_{0}.
\end{equation}
In accordance with (\ref{Lambda<0})  the parameters $\Lambda ,\Phi _{0}, V _{0},$ and $a$ should satisfy the inequality
 \begin{equation}\label{Lamda^tilda<0}
    \Lambda -\varkappa \left( \Phi _{0}^{2}/a+V_{0}\right) <0.
 \end{equation}

In general relativity the inequality (\ref{Lamda^tilda<0}), coming from the requirement of regularity, should replace the artificially imposed restriction
\begin{equation}\label{divfi = 0}
   \partial \phi ^{K}/\partial x^{K}=0,
\end{equation}
which people had been using for a long time to avoid the negative energy problem \cite{Bogolubov-Shirkov}.

\subsubsection{\label{subsubsection}Massive vector field}

If $V^{\prime }\neq 0$ the field equations (\ref{dFi'dx^I=... I=0,1})  allow to express $\phi _{0}$ and $\phi _{1}$ via $\Phi :$
\begin{equation}\label{fi_I=-m^-2...}
    \phi _{I}=-\frac{1}{V^{\prime }}\frac{\partial \Phi }{\partial x^{I}},\quad I=0,1.
\end{equation}
Substituting (\ref{fi_I=-m^-2...}) into (\ref{Fi=a(...) case a})  we get the following equation for $\Phi(x^{0},x^{1}):$
\begin{equation*}
    \frac{\partial ^{2}\Phi }{\left( \partial x^{0}\right) ^{2}}+d_{0}F^{\prime }\frac{\partial \Phi }{\partial x^{0}}-e^{-2F}\frac{\partial ^{2}\Phi }{\left( \partial x^{1}\right) ^{2}}+m^{2}\Phi =0.
\end{equation*}
The space coordinate $x^{1}$ is a cyclic variable, its conjugate momentum $p$ is the integral of motion, and the corresponding eigenfunction is
\begin{equation*}
    \Phi _{p}\left( x^{0},x^{1}\right) =\chi _{p}\left( x^{0}\right) e^{ipx^{1}}.
\end{equation*}
Function $\chi_{p}(x^{0})$  obeys the equation
\begin{equation}\label{hi''_p+...}
    \frac{d^{2}\chi _{p}}{\left( dx^{0}\right) ^{2}}+d_{0}F^{\prime }\frac{d\chi _{p}}{dx^{0}}+\left( p^{2}e^{-2F}+m^{2}\right) \chi _{p}=0.
\end{equation}

In flat space-time $F^{\prime }=0$ the solution is the sum of counter-propagating longitudinal plane waves
\begin{equation*}
    \Phi _{\left( p\right) }\left( x^{0},x^{1}\right) =C_{1}e^{i\left( px^{1}+E_{p}x^{0}\right) }+C_{2}e^{i\left( px^{1}-E_{p}x^{0}\right) },\quad F=F^{\prime }=0
\end{equation*}
with the relativistic spectrum of eigenvalues $E_{p}:$
\begin{equation*}
    E_{p}^{2}=p^{2}+m^{2}.
\end{equation*}
Both components $(\ref{fi_I=-m^-2...}) $ of the massive vector field are not zeroes.

Within short time periods $\Delta x^0 = x^0-x^{0*}\ll 1/F'$ the function $F(x^0)$ can be locally considered as a constant, depending on $x^0$ as a parameter. The local spectrum of eigenvalues is
\begin{equation*}
    E^{2}_{p}(x^0)=p^{2}e^{-2F(x^0)}+m^{2}.
\end{equation*}
Even if initially the field was ultrarelativistic, $p^2 \gg m^2,$ in the process of expansion the energy decreases, and the field finally becomes nonrelativistic.

In the de Sitter metric (\ref{de Sitter metric})  $dF=F^{\prime }dx^{0},$ $F^{\prime }=H/c=const$. After the substitution
\begin{equation}\label{z=e^-F}
    u=e^{-F}
\end{equation}
in the conformally equivalent metric
\begin{equation}\label{conformal metric}
    ds^{2}=e^{2F}\left( du^{2}-\sum_{I=1}^{d_{0}}\left( dx^{I}\right) ^{2}\right)
\end{equation}
the equation (\ref{hi''_p+...})  reduces to the Bessel equation
\begin{equation*}
    \frac{d^{2}\chi _{p}}{du^{2}}-\frac{d_{0}-1}{u}\frac{d\chi _{p}}{du}+\left( P^{2}+\frac{\mu^{2}}{u^{2}}\right) \chi _{p}=0.
\end{equation*}
There are two dimensionless parameters
\begin{equation}\label{dim-less parameters}
    P=\frac{p}{F^{\prime }}\text{ \ and }\mu =\frac{m}{F^{\prime }}.
\end{equation}
The solution is
\begin{equation}\label{hi_p....nu=1/2..}
    \chi _{p}=u^{\frac{1}{2}d_{0}}\left[ C_{1}\textmd{BesselY}_{\nu }\left( Pu\right) + C_{2}\textmd{BesselJ}_{\nu }\left( Pu\right) \right] ,\qquad \nu =\frac{1}{2}\sqrt{d_{0}^{2}-4\mu ^{2}}.
\end{equation}

If the mass $m$ is extremely small, namely
\begin{equation}\label{m/F'<<d_0}
    \mu \ll d_{0},
\end{equation}
then $m$ drops out of (\ref{hi_p....nu=1/2..}):
\begin{equation}\label{hi_p at m<<d_0F'}
    \chi _{p}=C_{1}u^{\frac{1}{2}d_{0}}\textmd{BesselY}_{\frac{d_{0}}{2}}\left( Pu\right) + C_{2}u^{\frac{1}{2}d_{0}}\textmd{BesselJ}_{\frac{d_{0}}{2}}\left( Pu\right) ,\quad \mu \ll d_{0}.
\end{equation}

The time variation of $\chi _{p}$ is damping oscillations of both addends in (\ref{hi_p at m<<d_0F'}). The argument $Pu=Pe^{-F}$ tends to zero at $x^{0}\rightarrow \infty .$ The second term in (\ref{hi_p at m<<d_0F'})  vanishes at $u\rightarrow 0,$ while the first one remains finite. For better visualization we rewrite the argument of Bessel functions as
\begin{equation}\label{w=}
    Pu=e^{-w},\qquad w=F^{\prime }x^{0}-\ln P.
\end{equation}
For $d_{0}=3$ the functions
\begin{equation}\label{red}
    e^{-\frac{1}{2}d_{0}w}\textmd{BesselY}_{\frac{d_{0}}{2}}\left( e^{-w}\right) =\allowbreak -\sqrt{\frac{2}{\pi }}\left( e^{-w}\sin e^{-w}+\cos e^{-w}\right)
\end{equation}
and
\begin{equation}\label{blue}
    e^{-\frac{1}{2}d_{0}w}\textmd{BesselJ}_{\frac{d_{0}}{2}}\left( e^{-w}\right) =\allowbreak -\sqrt{\frac{2}{\pi }}\left( e^{-w}\cos e^{-w}-\sin e^{-w}\right)
\end{equation}
are presented in Figure 1 by red and blue curves, respectively.
\begin{figure} \centering
\vspace{0cm}
    \hspace{0cm}
  \includegraphics{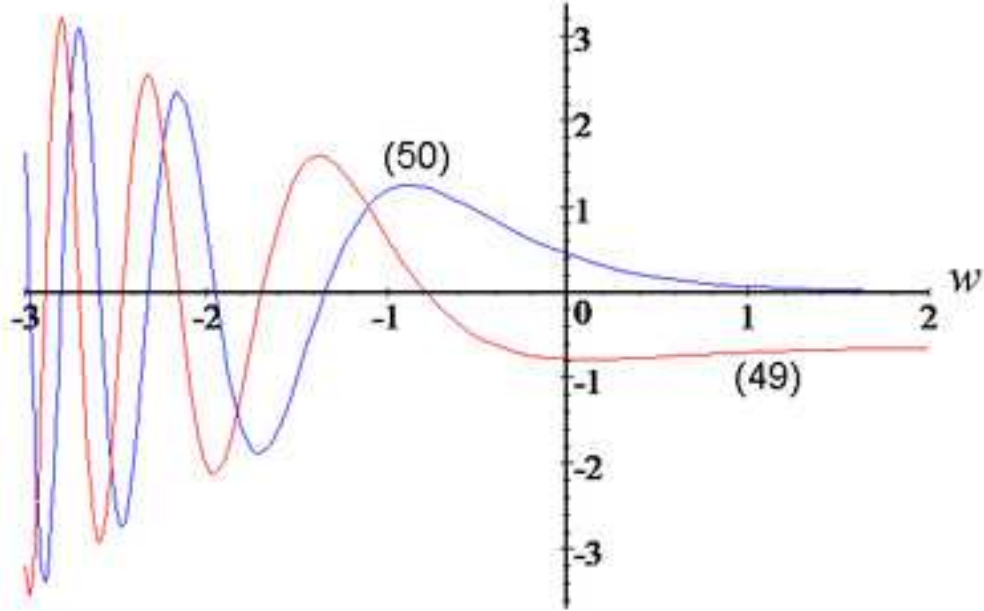}
  \caption{ Functions (\ref{red})  and (\ref{blue}). The second term in (\ref{hi_p at m<<d_0F'}) $-$ blue curve $-$ vanishes as $x^{0}\rightarrow \infty ,$ while the first one (red) remains finite.}\label{Fig1}
\end{figure}
The contribution of the zero mass vector field $\Phi $ to the energy-momentum tensor -- a possible source of the cosmological expansion -- remains finite at the late period of the Universe evolution.
	
In the opposite limit
\begin{equation}\label{m/F'>>d_0}
   \mu \gg d_{0}
\end{equation}
the index $\nu $ is large and imaginary. It is convenient to work with linear combinations of Bessel functions
\begin{equation*}
    H_{i\mu }^{\left( \pm \right) }\left( u\right) =\textmd{BesselJ}_{i\mu }\left( u\right) \pm i \textmd{BesselY}_{i\mu }\left( u\right) =\pm \frac{e^{\pm \mu \pi /2}}{\pi i}\int_{-\infty }^{\infty }e^{\pm iu\cosh t-i\mu t}dt.
\end{equation*}
and use their asymptotic at large indexes and arguments (tangent approximation \cite{Gradshtein}):
\begin{equation*}
    H_{i\mu }^{\left( \pm \right) }\left( u\right) =\frac{e^{\pm \mu \pi /2}}{\sqrt{2\pi }\left( u^{2}+\mu ^{2}\right) ^{1/4}}e^{\pm i\left[ \sqrt{u^{2}+\mu ^{2}}-\mu \text{Arsh}\left( \frac{\mu }{u}\right) -\frac{\pi }{4}\right] },\quad \mu \gg 1.
\end{equation*}

Asymptotic expression for $\chi _{p}$ \ at $\mu \gg 1:$
\begin{equation}\label{hi_p massive}
    \chi _{p}\left( u\right) =\frac{u^{d_{0}/2}}{\left( \mu ^{2}+u^{2}\right) ^{1/4}}\left( C_{1}e^{i\left[ \sqrt{u^{2}+\mu ^{2}}-\mu \text{Arsh}\left( \frac{\mu }{u}\right) -\frac{\pi }{4}\right] }+C_{2}e^{-i\left[ \sqrt{u^{2}+\mu ^{2}}-\mu \text{Arsh}\left( \frac{\mu }{u}\right) -\frac{\pi }{4}\right] }\right) ,\quad \mu \gg 1.
\end{equation}
Contrary to the case (\ref{m/F'<<d_0}), in the case of massive vector field (\ref{m/F'>>d_0})  function $\chi _{p}\left( u\right) $
(\ref{hi_p massive})  decreases with time and finally vanishes at $x^{0}\rightarrow \infty $ due to the factor $u^{d_{0}/2}\sim e^{-\frac{d_{0}}{2}F^{\prime }x^{0}}.$
	
In dimensional units the parameter $\mu =\frac{c}{H}\sqrt{\frac{V^{\prime }}{a}}=\frac{mc^{2}}{\hbar H}.$ With the today's value of the Hubble constant (\ref{H today})
 \begin{equation}\label{mu_today}
    \mu _{\text{today}}=\allowbreak 1.\,\allowbreak 3\times 10^{32}m=6.8\times 10^{37}\frac{m\text{ }}{m_{e}},
 \end{equation}
where $m$ is in eV, and $m_{e}$ is the electron mass. For all known massive particles $\mu _{\text{today}}$ is extremely large. The boundary between massive and massless vector fields $\mu \sim d_{0}$ corresponds to the masses $m\sim 10^{-32}$ eV. Practically today the case $\mu \ll d_{0}$ is indistinguishable from zero mass vector fields. The zero-mass vector field looks adequate for macroscopic description of the dark energy.
The principle difference between the two cases (massive fields vanish, and the massless ones survive at $x^{0}\rightarrow \infty $ ) approves the separation of dark relativistic sector into dark matter and dark energy.
	
In the hypothetical inflation period of the early Universe the Hubble constant was supposed to be extremely high (\ref{H inf}). In the stage of inflation the parameter $\mu $ \ should have been very small:
\begin{equation*}
    10^{-21}\frac{m}{m_{e}}<\mu _{\text{inf}}<10^{-15}\frac{m}{m_{e}}.
\end{equation*}
For the vector fields with masses up to $10^{15}m_{e}$ the parameter $\mu _{\text{inf}}$ is smaller than $d_{0},$ and the first term in (\ref{hi_p at m<<d_0F'}) does not vanish with time. The vector fields could have had all chances to survive the inflation epoch and contribute to the origin of today's expansion of the Universe.

\subsubsection{\label{subsubsection}Massless and massive fields acting together}
	
Massive and massless vector fields can be of different physical nature. In this case they are described by different functions. The Lagrangian has the form
\begin{eqnarray*}
  L =a_{e}\left( \phi _{e;M}^{M}\right) ^{2}-V_{0}+a_{m}\left[ \left( \phi _{m;M}^{M}\right) ^{2}-m^{2}\phi _{m}^{M}\phi _{m\text{ }M}\right] , \\
  V_{0} =V_{e}+V_{m}\left( 0\right) ,\quad m^{2}=V_{m}^{\prime }\left( 0\right) /a_{m}.
\end{eqnarray*}
Indexes $e$ and $m$ correspond to massless and massive fields, respectively. Constant $V_{m}\left( 0\right) $ is included into $\widetilde{\Lambda }$  via $V_{0},$ and the energy-momentum tensor of the massive field (\ref{T_IK    case a}) is
\begin{equation}
T_{m\text{ }IK}=a_{m}\left\{ g_{IK}\left( \phi _{m\text{ };M}^{M}\right) ^{2}+m^{2}\left( 2\phi _{m\text{ }I}\phi _{m\text{ }K}-\phi _{m\text{ }L}\phi _{m\text{ }}^{L}g_{IK}\right) \right\} .\label{T_mIK}
\end{equation}

The Einstein equations
\begin{equation}
   R_{IK}-\frac{1}{2}Rg_{IK}+\widetilde{\Lambda }g_{IK}=\varkappa T_{m\text{ }IK} \label{R_IK-...=T_mIK}
\end{equation}
    include the back reaction of both massive and massless vector fields. Their applicability is restricted by the condition that the second and higher derivatives of the potential $V_{m}\left( \phi _{m\text{ }L}\phi _{m\text{ }}^{L}\right) $ can be ignored. The back reaction of the massless field $\phi _{e}$ is included via $\widetilde{\Lambda }.$

    The energy-momentum tensor (\ref{T_mIK}) in the equations (\ref{R_IK-...=T_mIK})  contains only the massive field $\phi _{m}$, and therefore the index $m$ can be omitted. As far as the massive field does not violate the homogeneity and isotropy of the space,  $\phi _{m}$ does not depend on the space coordinates. Then in accordance with (\ref{fi_I=-m^-2...}) $\phi_1=0,$ and the Einstein equations (\ref{R_IK-...=T_mIK})  contain only the time-like component $\phi _{0}$ of the vector field:
\begin{eqnarray*}
	  \frac{1}{2}d_{0}\left( d_{0}-1\right) F^{\prime 2}+\widetilde{\Lambda } =\varkappa a\left[ \left( \frac{\partial \phi _{0}}{\partial x^{0}}+d_{0}F^{\prime }\phi _{0}\right) ^{2}+m^{2}\phi _{0}^{2}\right] ,  \\
  \left( d_{0}-1\right) F^{\prime \prime }+\frac{1}{2}d_{0}\left( d_{0}-1\right) F^{\prime 2}+\widetilde{\Lambda } =\varkappa a\left[ \left( \frac{\partial \phi _{0}}{\partial x^{0}}+d_{0}F^{\prime }\phi _{0}\right) ^{2}-m^{2}\phi _{0}^{2}\right] .
\end{eqnarray*}

 If there is no massive field, $\phi _{0}=0,$ the solution $F^{\prime }=\sqrt{-2\widetilde{\Lambda }/d_{0}\left( d_{0}-1\right) },$ $F^{\prime \prime }=0$ describes the expansion  without any acceleration or deceleration. The back reaction of a zero-mass longitudinal vector field reduces to the change of the constant rate of expansion.

 In the presence of a massive field $F^{\prime \prime }\neq 0:$ \begin{equation*}
      F^{\prime \prime }=-\frac{2a}{d_{0}-1}\varkappa m^{2}\phi _{0}^{2}=-\frac{2}{d_{0}-1}\varkappa V^{\prime }\left( 0\right) \phi _{0}^{2}.
\end{equation*}
Depending on the sign of the derivative $V^{\prime }\equiv \frac{dV\left( \phi _{L}\phi ^{L}\right) }{d\left( \phi _{L}\phi ^{L}\right) }$ at $\phi _{L}\phi ^{L}=0$ the expansion can go either with acceleration or deceleration. For instance, in the case of commonly used ``Mexican hat'' potential $V^{\prime }\left( 0\right) <0,$ and the expansion goes with acceleration.
	
The acceleration decreases and finally vanishes with time together with the massive vector field magnitude. The back reaction of the massive vector field can be ignored if the rate of expansion $F^{\prime }$ is close to its final constant value $F_{\infty }^{\prime }=\sqrt{-2\widetilde{\Lambda }/d_{0}\left( d_{0}-1\right) }:$
\begin{equation*}
    1-\frac{F^{\prime }}{F_{\infty }^{\prime }}\ll 1.
\end{equation*}

\subsection{\label{Subsection4.2}Transverse vector field}

We consider below a transverse vector field with antisymmetric covariant derivative $\left( a=0,\text{ }b=-c\right) $ in the background of the cosmological metric (\ref{de Sitter metric}). The field equations (\ref{Covar field eqs}) are
\begin{equation}\label{field eqs FIK}
bF_{IK}^{;K}=-V^{\prime }\phi _{I}/2.
\end{equation}
They are not gauge invariant because the vector field $\phi _{I}$ enters the equations directly, and not only via the covariant derivatives $F_{IK}.$ There is no freedom to impose Lorentz condition for massive vector fields.  Only if  $V^{\prime }=0$ the equations become gauge invariant.

The energy-momentum tensor (\ref{T_IK= general})  simplifies
 \begin{equation}\label{T_IK a=0, b=-c}
    T_{IK}=2b\left[ 4F_{\text{ \ }I}^{L}F_{LK}-g_{IK}F_{M}^{L}F_{\text{ }L}^{M}-2\left( F_{\text{ \ }K;L}^{L}\phi _{I}+F_{\text{ \ }I;L}^{L}\phi _{K}\right) \right] +g_{IK}V+2V^{\prime }\phi _{I}\phi _{K}
 \end{equation}
The vector $\phi _{I}$ of the transverse field has the space component $\phi _{1} $ directed along $x^{1}$ and depending on $x^{2}.$ The projection of the conjugate momentum $p_{2}\neq 0,$ and the equations (\ref{bDfi_0+...},\ref{b(Dfi_1-...})  with $\Phi =0$ reduce to $\left( F^{\prime \prime }=0,F^{\prime }=const\right) :$
\begin{eqnarray*}
  \widehat{D}\phi _{0}+\left[ 2\left( d_{0}-2\right) F^{\prime 2}+\frac{V^{\prime }}{b}\right] \phi _{0} =0 \\
 \widehat{D}\phi _{1}-2F^{\prime }\frac{\partial \phi _{1}}{\partial x^{0}}+\frac{V^{\prime }}{b}\phi _{1} =0.
\end{eqnarray*}
In the transverse vector field the components $\phi _{0}$ and $\phi _{1}$ are independent from one another. Imposing the Lorentz condition one can get only a partial solution.

In view of the translational symmetry along the space coordinates the momentum $p_{2}$ is the integral of motion. The vector field $\phi _{I}$ is a superposition of eigenfunctions $\phi _{I\left( p_{2}\right) }=\psi _{I\left( p_{2}\right) }\left( x^{0}\right) e^{ip_{2}x^{2}}.$ In the de Sitter metric (\ref{de Sitter metric})  $dF=F^{\prime }dx^{0}.$ The functions $\psi _{I\left( p_{2}\right) }\left( x^{0}\right) $ satisfy the equations
\begin{equation}\label{d^2fi_0/dF^2}
\frac{d^{2}\psi _{0}}{dF^{2}}+d_{0}\frac{d\psi _{0}}{dF}+\left[ P^{2}e^{-2F}+2\left( d_{0}-2\right) +\mu ^{2}\right] \psi _{0} =0
\end{equation}
\begin{equation}\label{d^2fi_1/dF^2+}
   \frac{d^{2}\psi _{1}}{dF^{2}}+\left( d_{0}-2\right) \frac{d\psi _{1}}{dF}+\left[ P^{2}e^{-2F}+\mu ^{2}\right] \psi _{1} =0
\end{equation}
with the same two dimensionless parameters $\frac{p_{2}}{F^{\prime }}=P,\quad \frac{m^{2}}{F^{\prime 2}}=\mu ^{2}$ as in the longitudinal case (\ref{dim-less parameters}), except that now $m^{2}=\frac{V^{\prime }}{b}.$ The vector field is nonrelativistic if  $Pe^{-F}\ll \mu ,$ and it is ultrarelativistic until $Pe^{-F}\gg \mu .$

After the substitution (\ref{z=e^-F}) in the conformally equivalent metric (\ref{conformal metric})  the eqs. (\ref{d^2fi_0/dF^2},\ref{d^2fi_1/dF^2+}) acquire the form of Bessel equations
\begin{equation*}
\frac{d^{2}\psi _{0}}{du^{2}}-\frac{d_{0}-1}{u}\frac{d\psi _{0}}{du}+\left( P^{2}+\frac{2d_{0}-4+\mu ^{2}}{u^{2}}\right) \psi _{0} =0
\end{equation*}
\begin{equation*}
\frac{d^{2}\psi _{1}}{du^{2}}-\frac{d_{0}-3}{u}\frac{d\psi _{1}}{du}+\left( P^{2}+\frac{\mu ^{2}}{u^{2}}\right) \psi _{1} =0.
\end{equation*}
The solutions are
\begin{equation}\label{psi_0(u)=}
    \psi _{0}\left( u\right) =u^{\frac{d_{0}}{2}}\left[ C_{01}\textmd{BesselY}_{\nu _{0}}\left( Pu\right) +\allowbreak C_{02}\textmd{BesselJ}_{\nu _{0}}\left( Pu\right) \right]
\end{equation}
\begin{equation}\label{psi_1(u)=}
    \psi _{1}\left( u\right) =u^{\frac{d_{0}}{2}-1}\left[ C_{11}\textmd{BesselY}_{\nu _{1}}\left( Pu\right) +\allowbreak C_{12}\textmd{BesselJ}_{\nu _{1}}\left( Pu\right) \right] ,
\end{equation}
where $\nu _{0}=\frac{1}{2}\sqrt{\left( d_{0}-4\right) ^{2}-4\mu ^{2}},$ $\ \nu _{1}=\sqrt{\left( 1-\frac{d_{0}}{2}\right) ^{2}-\mu ^{2}},$ and $C_{ik}$ are integration constants, $i=0,1$, $k=1,2$.
	
At $\mu \gg d_{0}$\ the component $\psi _{0}\left( u\right) $ (\ref{psi_0(u)=})  of the transverse field is the same as (\ref{hi_p....nu=1/2..})  for the longitudinal field. There is no need to repeat the same statements as in the end of the previous section concerning the asymptotic behavior of massive transverse fields at $\mu \gg d_{0}.$ Both functions (\ref{psi_0(u)=}) and (\ref{psi_1(u)=})  vanish at $x^{0}\rightarrow \infty $ if $\mu \gg d_{0}>2.$
	
\subsubsection{\label{subSubsection4.2}Electromagnetic field in the expanding Universe}

If $\mu =0$ the energy-momentum tensor (\ref{T_IK a=0, b=-c})  reduces to the one of the electromagnetic field:
\begin{equation*}
   T_{IK}=2b\left( 4F_{\text{ \ }I}^{L}F_{LK}-g_{IK}F_{M}^{L}F_{\text{ }L}^{M}\right) +g_{IK}V_{0}.
\end{equation*}
The equations (\ref{field eqs FIK}) become gauge invariant, and the vector field $\phi _{I}$\  describes the electromagnetic field in the expanding Universe in the absence of electric charges.
The two projections of the field, a time-like $\phi _{0},$ and a space-like $\phi _{1},$ are independent from one another. Their temporal evolution is different:
\begin{equation}\label{psi_0(u)=...mu=0}
   \psi _{0}\left( u\right)  =C_{01}u^{\frac{1}{2}d_{0}}\textmd{BesselJ}_{|2-\frac{1}{2}d_{0}|}\left( Pu\right) +\allowbreak C_{02}u^{\frac{1}{2}d_{0}}\textmd{BesselY}_{|2-\frac{1}{2}d_{0}|}\left( Pu\right) ,
\end{equation}
\begin{equation}\label{psi_1(u)=...mu=0}
    \psi _{1}\left( u\right)  =C_{11}u^{\frac{1}{2}d_{0}-1}\textmd{BesselJ}_{\frac{1}{2}d_{0}-1}\left( Pu\right) +\allowbreak C_{12}u^{\frac{1}{2}d_{0}-1}\textmd{BesselY}_{\frac{1}{2}d_{0}-1}\left( Pu\right) ,\quad u=e^{-F}.
\end{equation}
In the case of three space dimensions $d_{0}=3$
\begin{equation}\label{psi0(u)}
    \psi _{0}\left( u\right)=\allowbreak \sqrt{\frac{2}{\pi P}}u\left( C_{01}\cos Pu+C_{02}\sin Pu\right)
\end{equation}
\begin{equation}\label{psi1(u)}
    \psi _{1}\left( u\right) =\allowbreak \sqrt{\frac{2}{\pi P}}\left( C_{11}\cos Pu+C_{12}\sin Pu\right)
\end{equation}
The component $\phi _{0}$ (\ref{psi0(u)})  oscillates with a decreasing frequency and vanishes with time due to the factor $u=e^{-F}$ (blue curves in Figures 2a and 2b for terms with $\textmd{BesselY}$ and $\textmd{BesselJ}$, respectively). The space projection $\phi _{1}$ (\ref{psi1(u)})  oscillates with a constant amplitude and decreasing frequency (red curves in Figures 2a and 2b).
\begin{figure} \centering
    \vspace{0cm}
    \hspace{0cm}
    \includegraphics{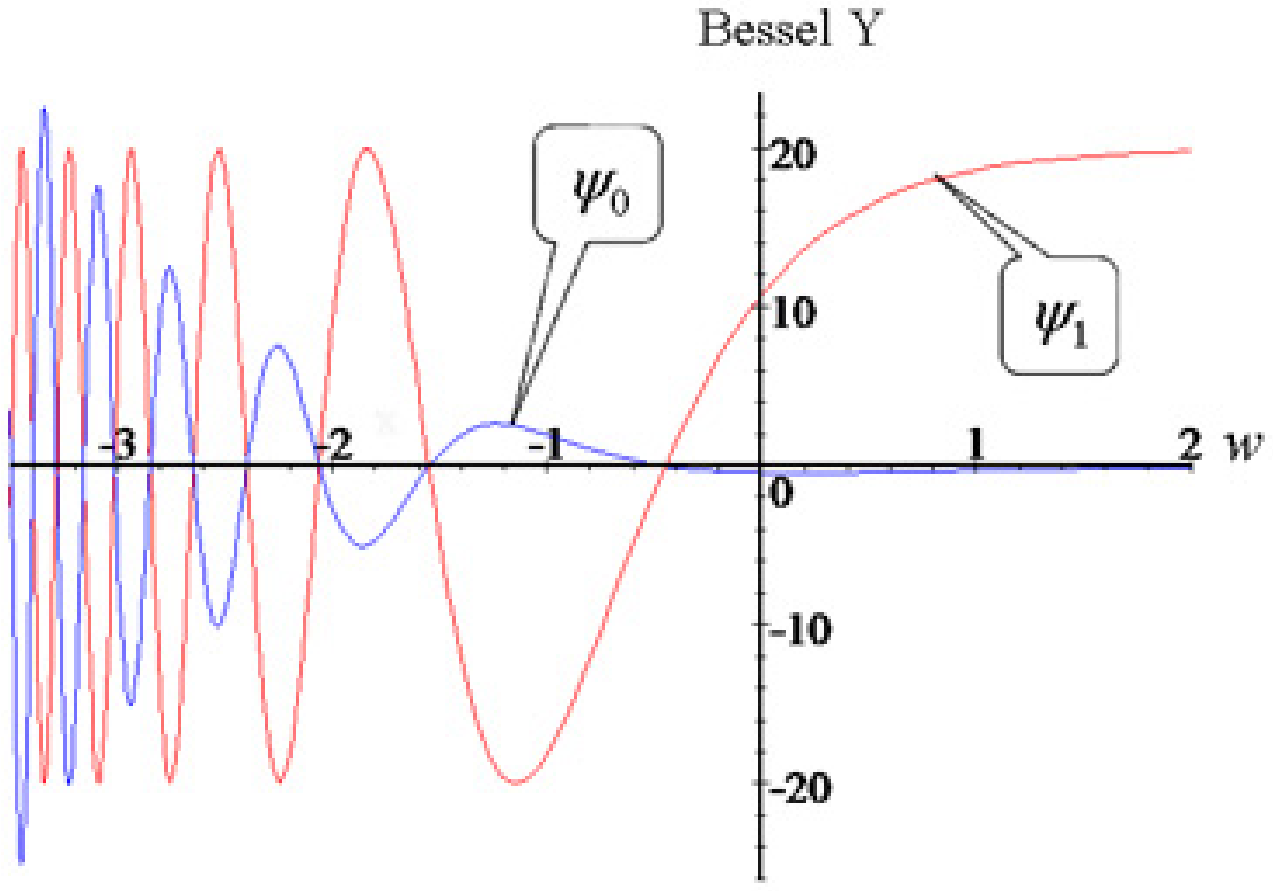}   \\  
    \includegraphics{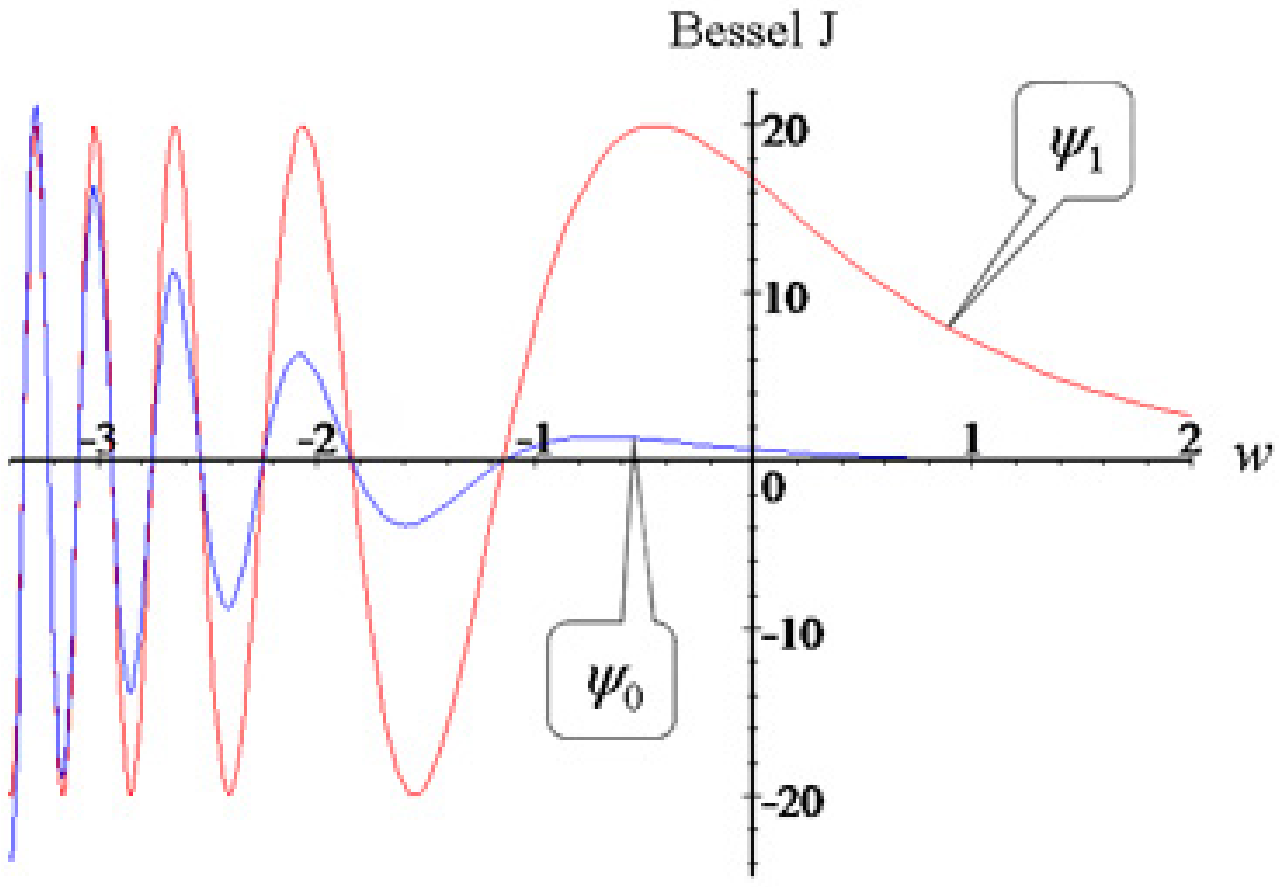}
    \caption{\label{Fig2} Upper: Transverse vector field in the background of the de Sitter metric. $d_{0}=3$. Projections $\psi _{0}$ (blue) and $\psi _{1}$ (red), both $\sim \textmd{BesselY}$, as functions of $w$ (\ref{w=}).   \\
        Lower: Transverse vector field in the background of the de Sitter metric. $d_{0}=3$. Projections $\psi _{0}$ (blue) and $\psi _{1}$ (red), both $\sim \textmd{BesselJ}$, as functions of $w$ (\ref{w=}).}
\end{figure}
In the conformally equivalent metric (\ref{conformal metric})  at $d_{0}=3$ the space projection $\phi _{1}$ is an ordinary plane wave. Without expansion ($F^{\prime }=0)$ there would be no difference between the two projections (\ref{psi0(u)})  and (\ref{psi1(u)})  of the transverse vector field.
	
If the number of space dimensions $d_{0}>3$ the amplitude of oscillations decreases with time, and if $d_{0}<3$ -- grows. The behavior of the space projections $\phi _{1}\sim \textmd{BesselJ}$ in (\ref{psi_1(u)=...mu=0}) are presented in Figure 3a for $d_{0}=4$ (brown curve) and $d_{0}=2$ (blue curve). Similar curves for $\phi _{1}\sim \textmd{BesselY}$ are presented in Figure 3b.
\begin{figure} \centering
    \vspace{0cm}
    \hspace{0cm}
    \includegraphics{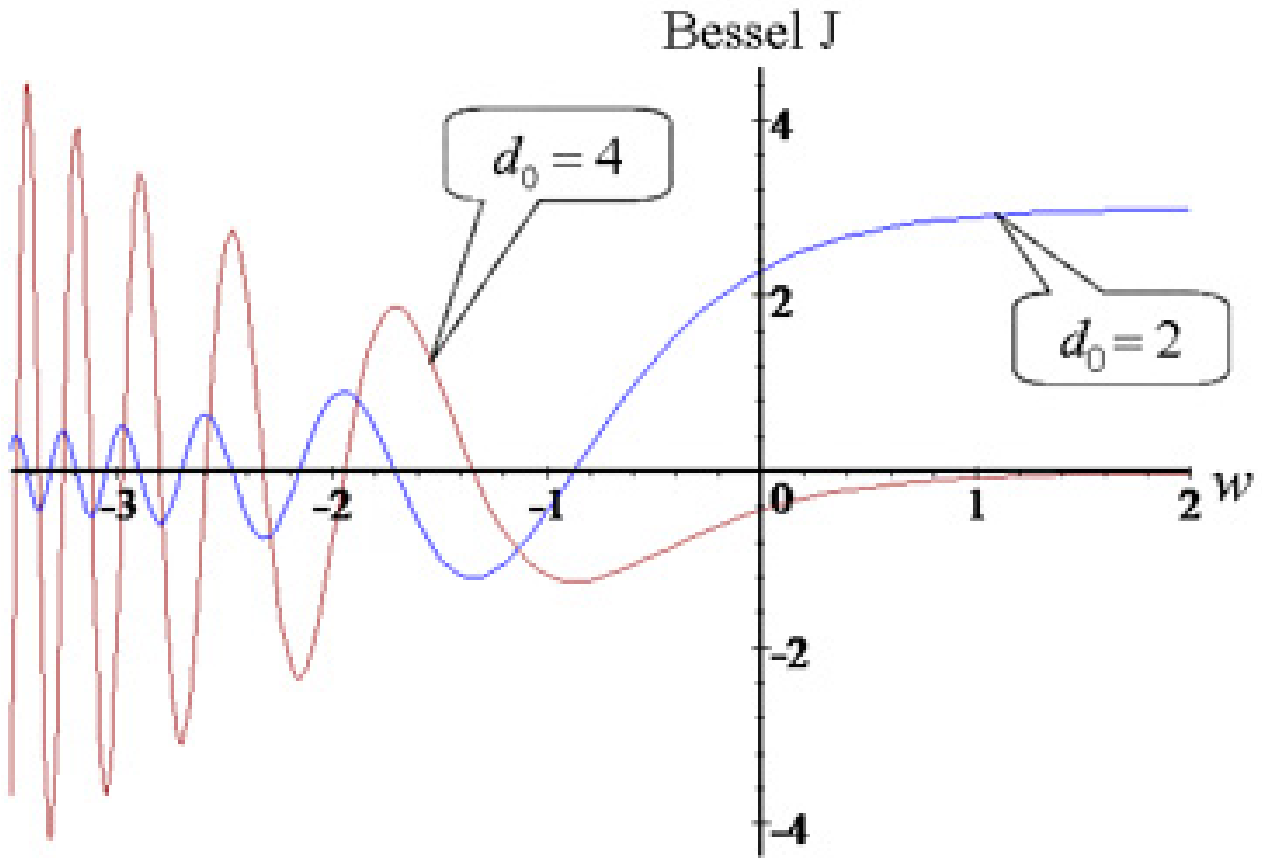}   \\  
    \includegraphics{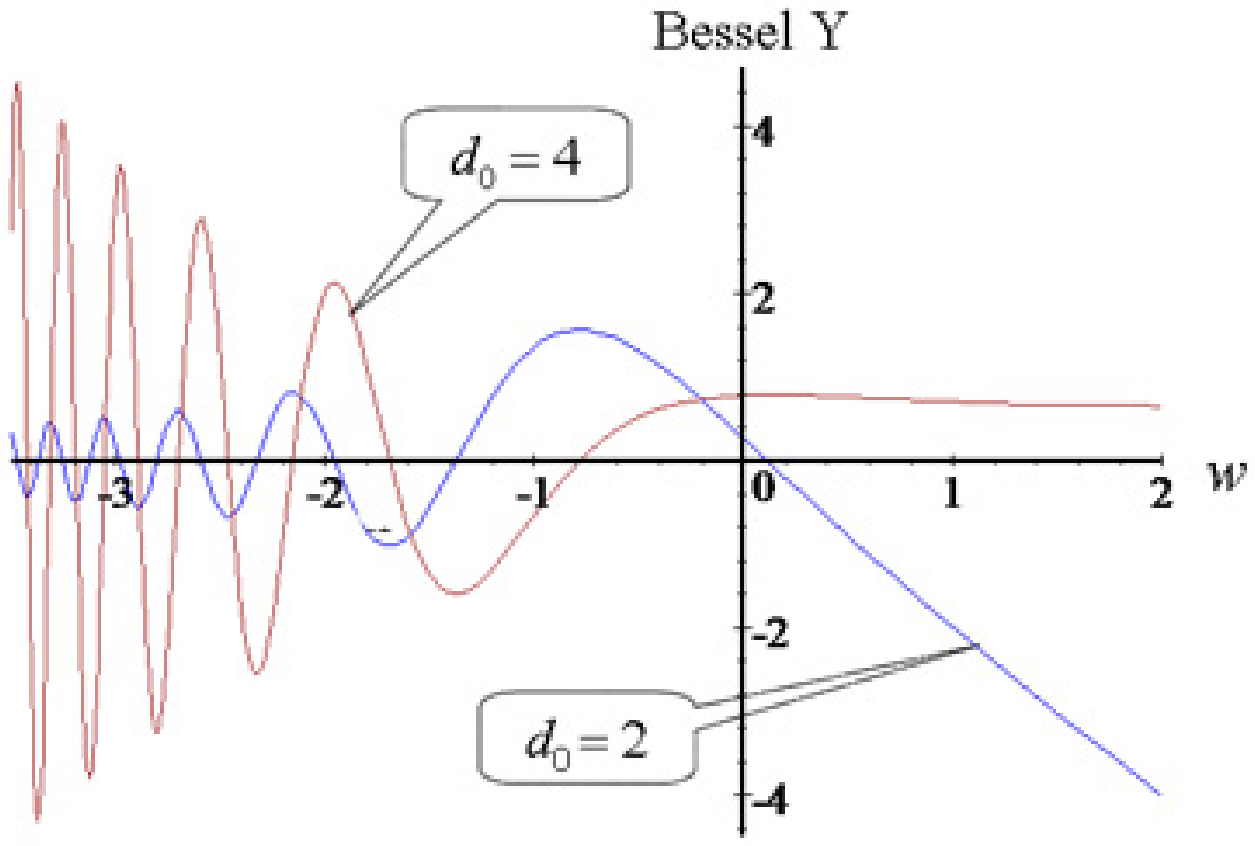}
    \caption{\label{Fig3} Upper: The behavior of the space projections $\phi _{1}\sim \textmd{BesselJ}$ in (\ref{psi_1(u)=...mu=0})  for $d_{0}=4$ (brown curve) and $d_{0}=2$ (blue curve).   \\
        Lower: The behavior of the space projections $\phi _{1}\sim \textmd{BesselY}$ in (\ref{psi_1(u)=...mu=0})  for $d_{0}=4$ (brown curve) and $d_{0}=2$ (blue curve). }
\end{figure}
The frequency of oscillations decreases with time exponentially.

In the two particular cases $( a\neq 0, b=c=0) $ and  $( a=0, b=-c), $ considered above, the vector fields in the uniform bulk with $d_{0}>3$ decrease in time and thus can not initiate a phase transition resulting in the formation of a brane. For $d_{0}=2 $  $\phi _{1}\sim \textmd{BesselY}$ grows with time (see Figure 3b), but it has nothing to do with the brane formation. The question, if for some other values of  $ a, b, c $ and  $d_{0}>3 $ the vector fields can grow with time, remains open.

During short time intervals $\left( x^{0\ast },x^{0}\right) $ (as compared with the life time of the Universe $\left| x^{0}-x^{0\ast }\right| \ll 1/F^{\prime }$) the exponent in (\ref{z=e^-F})  can be expanded in powers of $x^{0}-x^{0\ast }.$ It is convenient to choose the scale of the axis $ x^{2}$ so that $F\left( x^{0\ast }\right) =0.$ To the accuracy of $\left( x^{0}-x^{0\ast }\right) ^{2}$
\begin{equation*}
    Pu=\frac{p_{2}}{F^{\prime }}-\widetilde{p_{2}}\left( x^{0}-x^{0\ast }\right) ,\text{ \ \ \ \ }\widetilde{p_{2}}\left( x^{0}\right) =p_{2}\left[ 1-\frac{1}{2}\left( F^{\prime }-\frac{F^{\prime \prime }}{F^{\prime }}\right) \left( x^{0}-x^{0\ast }\right) \right] .
\end{equation*}
Let the frequency of the vector field be$\ \omega =\frac{p_{2}c}{\hbar }$ at $x^{0}=x^{0\ast }.$ Due to the expansion of the Universe the frequency at a later moment $x^{0}$ becomes shifted to $\widetilde{\omega }=\frac{\widetilde{p_{2}}c}{\hbar }$.\allowbreak\ In the de Sitter metric (\ref{de Sitter metric})  $F^{\prime \prime }=0,F^{\prime }=H/c,$ and
\begin{equation*}
    \frac{\widetilde{\omega }}{\omega }=1-\frac{1}{2}H\left( t-t^{\ast }\right).
\end{equation*}
Here $t^{\ast }$ is the moment of emission, and $t$ is the moment of observation.

\section{\label{Summary}Summary}
	
Among the three-parametric variety of vector fields ($a,b,$ and $c$ are arbitrary constants) the two particular cases are analyzed here in detail. Namely, longitudinal fields ($b=c=0),$ and transverse (including electromagnetic) fields $\left( a=0,\text{ }b=-c\right).$
	The longitudinal zero-mass vector field $\phi _{I}$ with the Lagrangian (\ref{Lagrangian case a})  pretends to be an adequate tool for macroscopic description of dark energy as the major source of the observable expansion of the Universe. The energy-momentum tensor of a zero-mass vector field (\ref{T_IK case a m=0})  enters the Einstein equations (\ref{Einstein equations General})  as an addition to the cosmological constant. The solution -- de Sitter metric -- describes the homogeneous and isotropic expansion at a constant rate. The zero-mass vector field does not vanish with time in the process of expansion. It tends to a constant value at $t\rightarrow \infty ,$ see the red curve in the Figure 1.
	
On the contrary, massive vector fields decrease with time, $\phi _{I}\rightarrow 0$ at $t\rightarrow \infty ,$ provided that the expansion goes at the modern rate. Nevertheless, massive fields could survive the epoch of inflation. Though their amplitude is falling down, it is possible that they are still contributing to the evolution of the Universe,
making the expansion accelerated.

According to modern observations among the staff of the Universe the biggest 'slice of the cake' - 72 $^{0}/_{0}$ - corresponds to dark energy, 23 $^{0}/_{0}$ is dark matter, and only 4.6 $^{0}/_{0}$ comes from atoms \cite{NASA}. Vector fields of different physical nature can be characterized each  by its own set of the parameters  $a, b,$ and $c$. It is interesting, to what extent the appropriate choice of the parameters can fit this distribution? But it is a different story.
	
The general analysis of vector fields in the background of cosmological metric and the two particular cases considered analytically in detail confirm that the abilities of the Einstein's theory of general relativity to guess the riddles of cosmological observations are not yet exhausted.
	
\section{\label{acknowledgement}Acknowledgement}

I am grateful to K.A.Bronnikov and V.I.Marchenko for discussions.


\end{document}